\documentclass[final,3p,times,twocolumn]{elsarticle}

\usepackage{graphicx}

\usepackage{amssymb}
\usepackage{amsthm}
\usepackage{lmodern}

\usepackage{color}
\usepackage{verbatim}

\def\STDgas{\rm C_2F_4H_2}
\def\ECOgas{\rm C_3F_4H_2}
\def\SF6{\rm SF_6}
\def\CF3I{\rm CF_3I}
\def\CO2{\rm CO_2}

\biboptions{square}

\journal{Nuclear Instruments and Methods in Physics Research}
\begin{document}

\begin{frontmatter}



\title{Study of the ecological gas for MRPCs}


\author[1,2]{Yongwook Baek\corref{corr}}
\cortext[corr]{Correspoding author.}
\ead{yong.wook.baek@cern.ch}
\author[1,3]{Dowon Kim}
\author[1,2,4]{M.C.S. Williams}
\address[1]{Gangneung-Wonju National University, South Korea}
\address[2]{CERN, Switzerland}
\address[3]{ICSC World Laboratory, Geneva, Switzerland}
\address[4]{INFN-Bologna, Italy}

\begin{abstract}
The Multigap Resistive Plate Chamber (MRPC) is a gaseous detector; the performance depends very much on the gas mixture as well as the design. 
MRPCs are used as a timing device in several collider experiments and cosmic ray experiments thanks to the excellent timing performance. The typical gas mixtures of RPC-type detectors at current experiments are based on the gases $\STDgas$ and $\SF6$. These gases have very high Global Warming Potential (GWP) values of 1430 and 23900 respectively. 

The present contribution has been performed as a part of efforts to reduce the amount of greenhouse gases used in high energy experiments.  The performance of MRPC has been measured with two different gas mixtures; $\STDgas$ based gas mixtures and the ecological $\ECOgas$ (HFO-1234ze). A small MRPC was used for the tests.  It has an sensitive area of 20 $\times$ 20\,$\rm cm^2$; it was been built with  6 gaps of 220\,$\mu$m. 

In normal operation, the strong space charge created within the gas avalanche limits the avalanche's growth. $\SF6$ plays an important part in the process due to its high attachment coefficient at low electric fields.   It is thus necessary to find another gas that has a similar attachment coefficient.  $\CF3I$ is a possible candidate.  Tests were performed with this gas added to $\ECOgas$. 
\end{abstract}

\begin{keyword}
LHC, ALICE-TOF, MRPC, eco-friendly gas


\end{keyword}

\end{frontmatter}


\section{Motivation}
The advantage of using the Resistive Plate Chamber (RPC) is due to its low cost and high detection efficiency.  Most RPC-type detectors have been operating with gas mixtures containing R134a ($\STDgas$) and $\SF6$, that have a high Global Warning Potential (GWP). 
As a way of reducing an amount of harmful gases emitted to the atmosphere, closed loop gas systems have been introduced, however the construction cost is not negligible.
A better way is to replace these high GWP gases with more ecological gas mixtures. 

Searching for new ecological gases has been carried out by various groups~\cite{eco:helium,eco:RPC}, this study continues this investigation.  The ecological gas $\ECOgas$ has been considered as a substitute for $\STDgas$, and gas mixtures with the following gases, $\CO2$, $\SF6$ and $\CF3I$, have been tested. Especially, $\CF3I$ is introduced as a possible candidate to substitute for $\SF6$.

\section{MRPC}
We used a MRPC that has an active area of 20 $\times$ 20\,$\rm cm^2$. It consists of 24 pickup strips of 0.7 $\times$ 20.5\,$\rm cm^2$ with 6 gas gaps of 220\,$\mu$m.  The resistive plates were 280\,$\mu$m thick soda lime float glass.  The measured time resolution in previous tests with gas mixture, $\STDgas/\SF6$ (95\%/5\%)~\cite{FORSTER2016182} is about 80\,ps at 15\,kV operating voltage. 

The ultrafast NINO amplifier-discriminator~\cite{Anghinolfi:818554}  is used to readout signals from MRPC. This ASIC has been developed for  ALICE TOF detector at LHC experiment~\cite{Aamodt:2008zz}.  A differential signal is derived from anode and cathode readout strips which is used as input to the NINO card. The width of the output signal from the NINO is related on the amplitude of the input signal through a time-over-threshold (TOT) technique.  
\begin{figure}[t]
\centering
\includegraphics[width=0.4\textwidth]{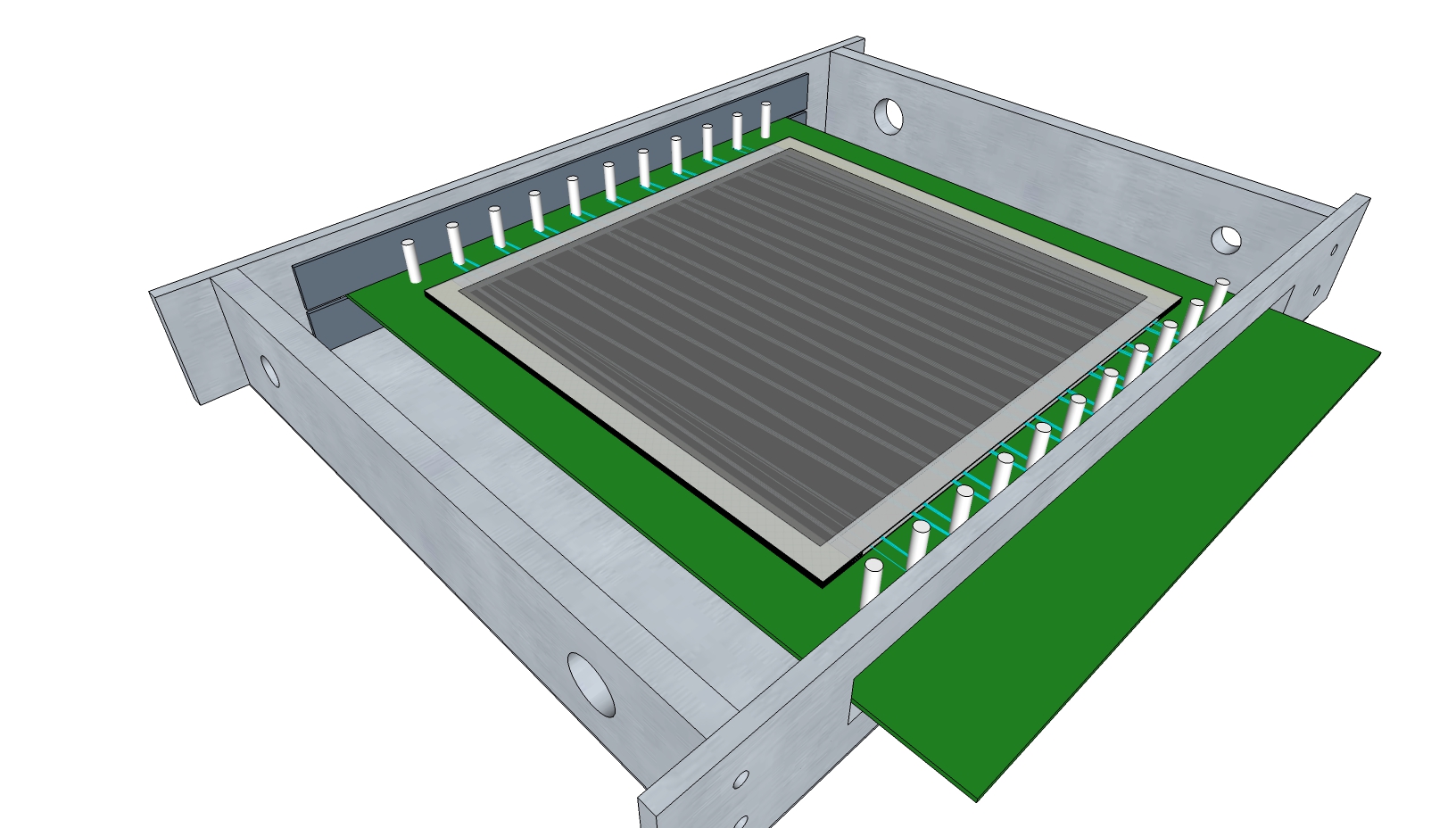}
\caption{Schematic view of the small MRPC used for the beam test. It has  a 20 $\times$ 20\,$\rm cm^2$ active area with 6 gaps of 220\,$\mu$m.}
\label{smallMRPC}
\end{figure}

The signals at both ends of a strip of the small MRPC have been readout by NINO cards. The threshold control voltage was set at 160\,mV; this threshold setting corresponds to the discriminator being set to fire for 40\,fC signals.  The NINO's LVDS signal has been connected to two CAEN V1290 TDCs which have a time resolution of 30\,ps for the two edges of the LVDS pulse. 

\section{Experimental layout}
\begin{figure}[th]
\centering
\includegraphics[width=0.47\textwidth]{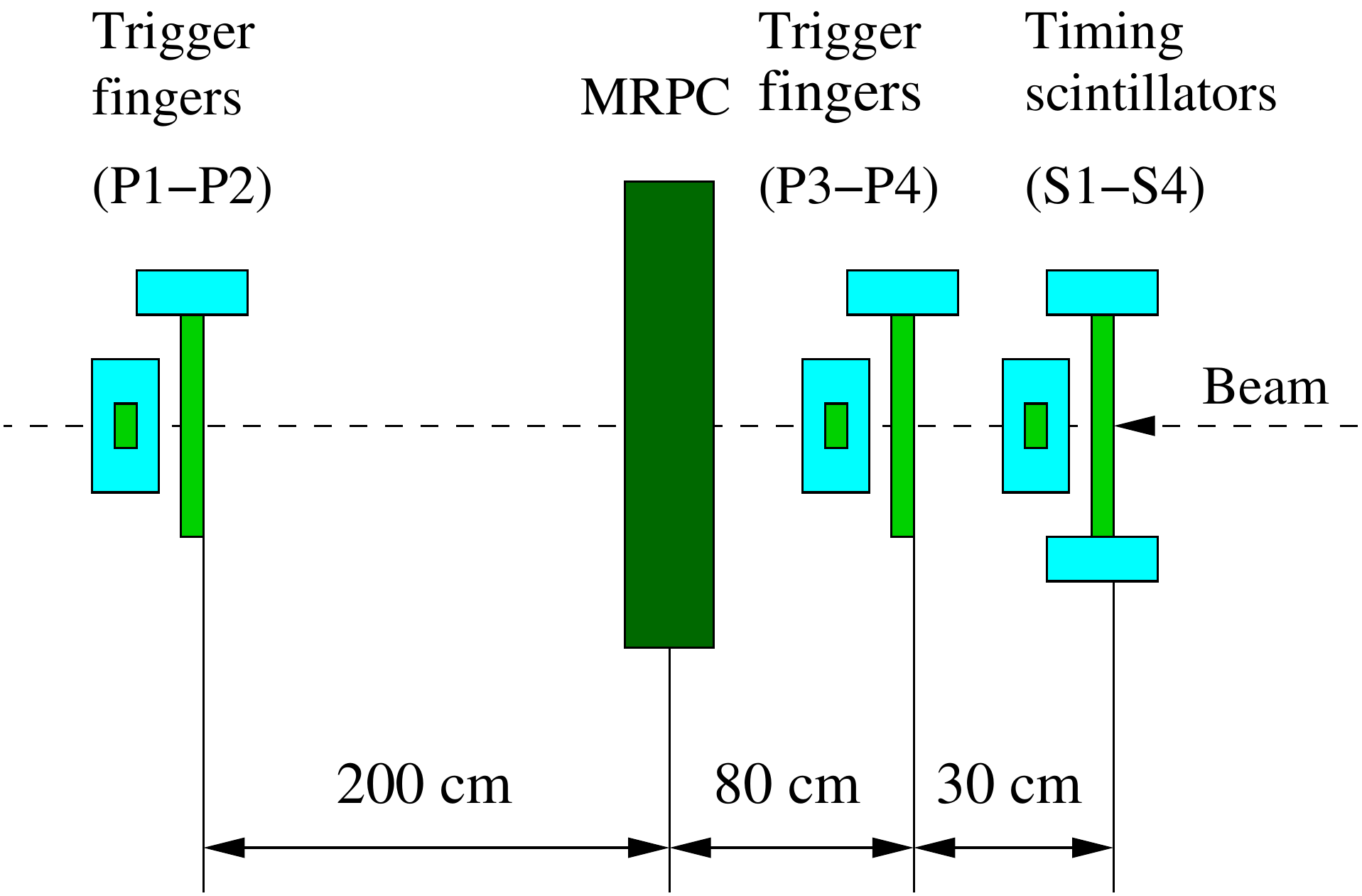}
\caption{Setup at T10 beam facility.}
\label{T10setup}
\end{figure}

T10 is a test beam in the East hall; it provides a pion beam with a momentum up to 6\,GeV/c. The setup at T10 is shown in figure~\ref{T10setup}.  Two sets of trigger scintillators with active areas of 2 $\times$ 2\,$\rm cm^2$  for P1-P2 and 1 $\times$ 1\,$\rm cm^2$ for P3-P4. A trigger signal is created from the coincidence of these scintillators. The accurate time information of triggered event is provided by two fast scintillator bars (2 $\times$ 2 $\times$ 10\,$\rm cm^3$); each bar is read out with two photomultipliers.  These are shown as S1-S4 in figure~\ref{T10setup}.

\section{Results}
\subsection{Event selection}
The preliminary stage of event selection has been done by accepting events within $\pm 3\sigma$ from the mean value of the time difference distribution from two trigger scintillators; (S1+S2)/2 - (S3+S4)/2.  The mean time, (S1+S2+S3+S4)/4, is used as reference time.  Its precision is estimated from the time difference distribution of two timing  scintillators giving a time resolution of the reference time of 47\,ps.  When we quote the time resolution of the MRPC this 47\,ps is subtracted in quadrature.    The data was collected in two test beam periods with the beam intensity set at various low flux. The flux is estimated from the coincidence of the scintillators S1-S4 with a sensitive area 2 $\times $ 2\,cm$^2$ and the beam spill period of  350\,ms.

\subsection{Efficiency}
\begin{figure}[t]
\centering\includegraphics[width=0.5\textwidth]{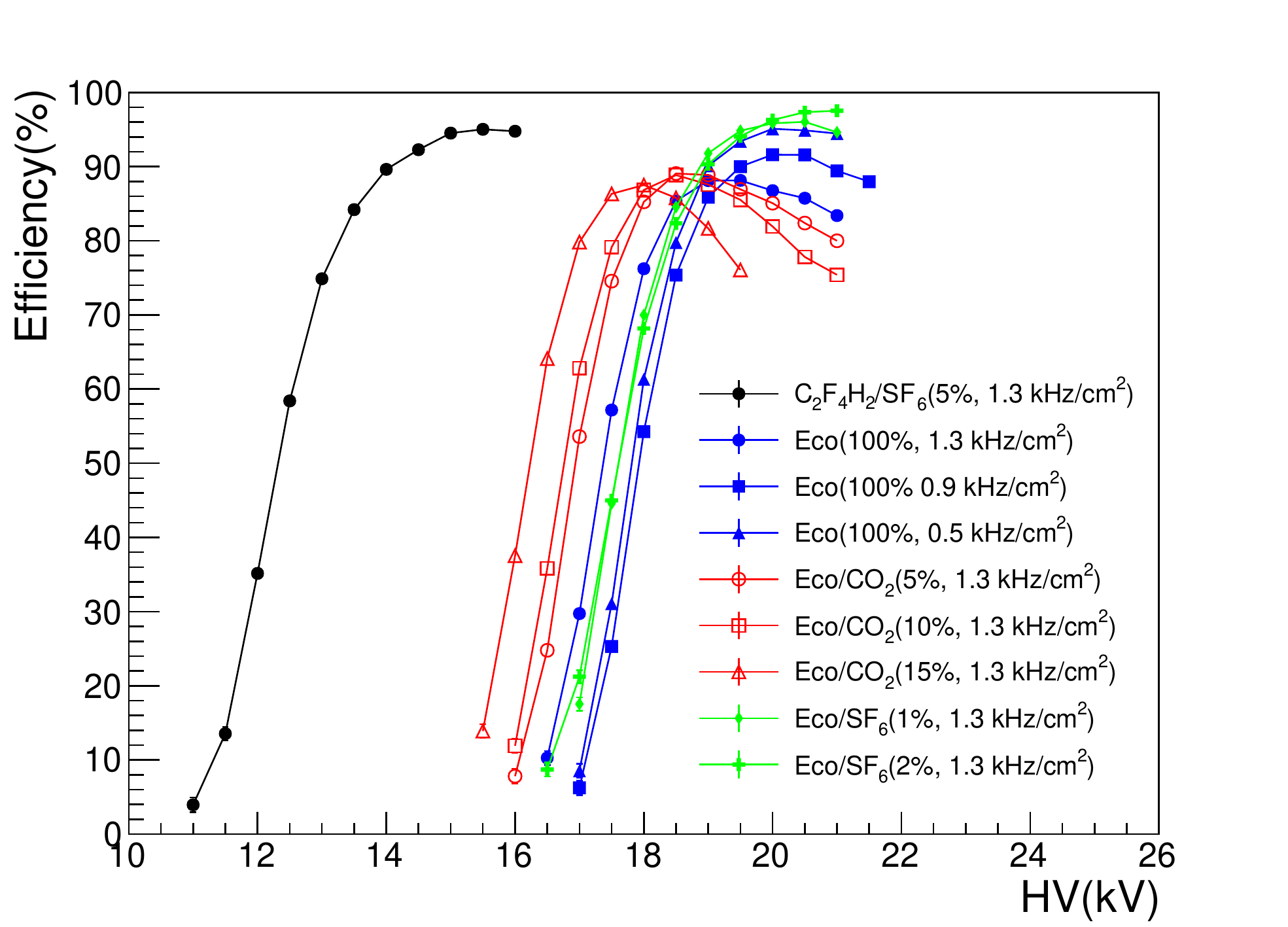}
\caption{Efficiencies measured with various gas mixtures at 1.3 kHz/cm$^{2}$ of particle flux and 100\% $\ECOgas$ is tested at additional particle flux, 0.5 and 0.9 kHz/cm$^2$.The error bars are contained within the size of the symbols.}
\label{fig:efficiency1300}
\centering\includegraphics[width=0.5\textwidth]{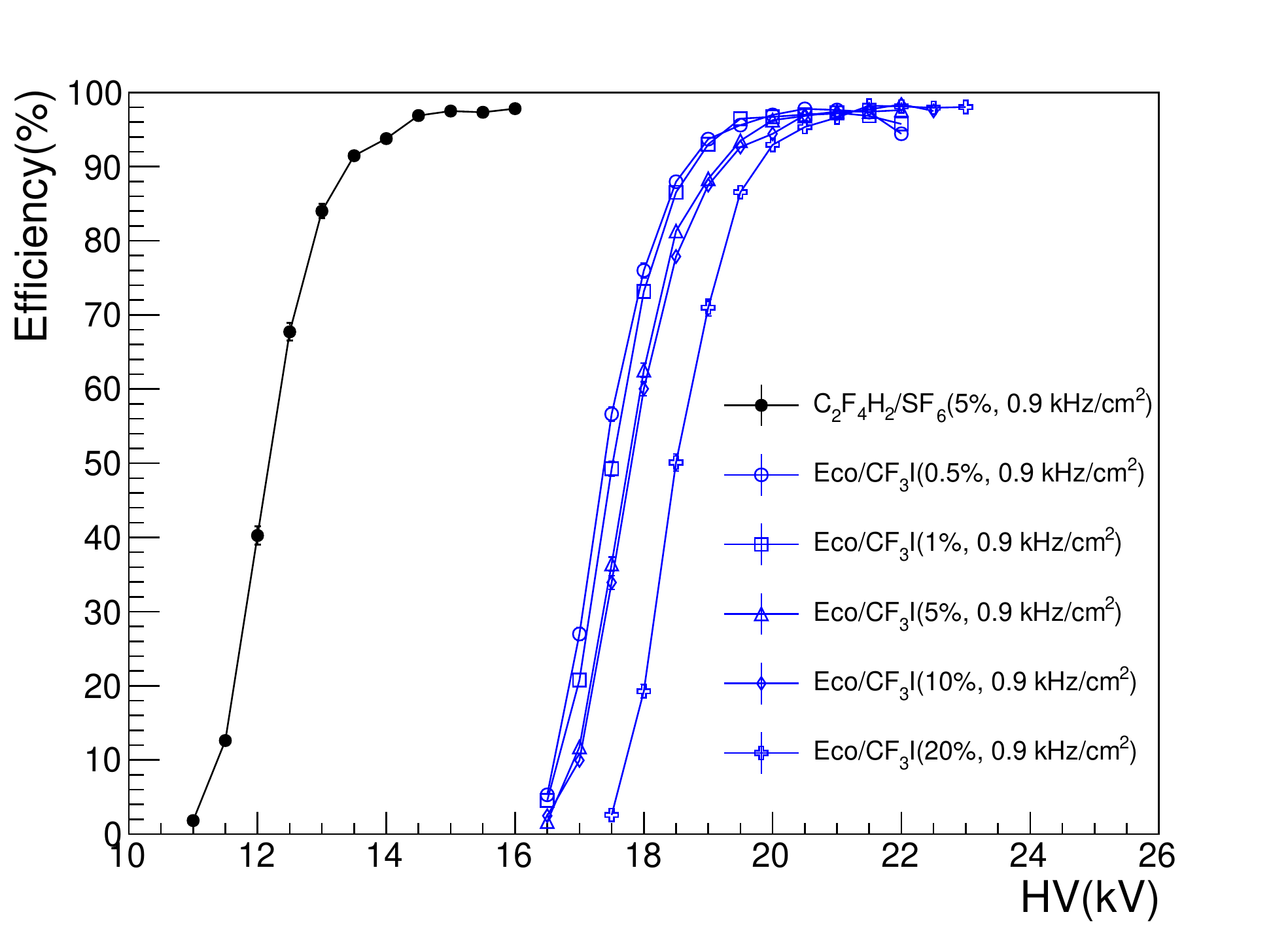}
\caption{Efficiencies measured with various gas mixtures that includes $\CF3I$ mixed with $\ECOgas$.}
\label{fig:efficiency900}
\end{figure}

The chamber efficiency is defined as a ratio between the number of events detected by MRPC and the number of triggered events.   The figures \ref{fig:efficiency1300}  and \ref{fig:efficiency900} show efficiencies as a function of applied voltages, obtained with various gas mixtures and particle flux. 

In the first period, the default particle flux was 1.3\,kHz/cm$^2$, and two additional flux, 0.5 and 0.9\,kHz/cm$^2$ are used for the 100\% $\ECOgas$ to check behaviour of the MRPC at various rates Basically, the operating voltage of $\ECOgas$ mixtures need an increase of voltage by 4\,kV more than the $\STDgas/\SF6$ gas mixture. 
The efficiencies obtained for the efficiency plateau  are 95\% for $\STDgas/\SF6$ and 88\% for $\ECOgas$ at the same particle flux of 1.3\,kHz/cm$^{2}$.  At lower flux for  pure $\ECOgas$ the efficiency increases and the plateau becomes longer, and in case of 0.5\,kHz/cm$^2$  we obtained the similar result as the one of $\STDgas/\SF6$. 

Adding $\CO2$ to the $\ECOgas$ and increasing the ratio up to 15\%, the operating voltage reduced by 1\,kV, however the efficiency does not improved and the efficiency plateaus shortened.
$\SF6$ is known as a very electronegative gas together with the highest known GWP of 23900. Adding this to $\ECOgas$ the efficiency increases and reaches 98\%. The plateaux lengthen even with a small amount of $\SF6$.

 In the second beam period, a new gas $\CF3I$ has been tested, which is a part of the research for finding a new gas in order to replace high GWP gas $\SF6$. Adding this new gas to the ecological gas increases the efficiency.   Increasing the ratio of $\CF3I$ gas needs higher operating voltage. The plateaux are improved by becoming longer, shown in figure~\ref{fig:efficiency900}.  The efficiencies are 98\% for both gas mixtures;  $\STDgas/\SF6$ and $\ECOgas/\CF3I$.

\subsection{Time resolution}
\begin{figure}[!p]
\centering\includegraphics[width=0.5\textwidth]{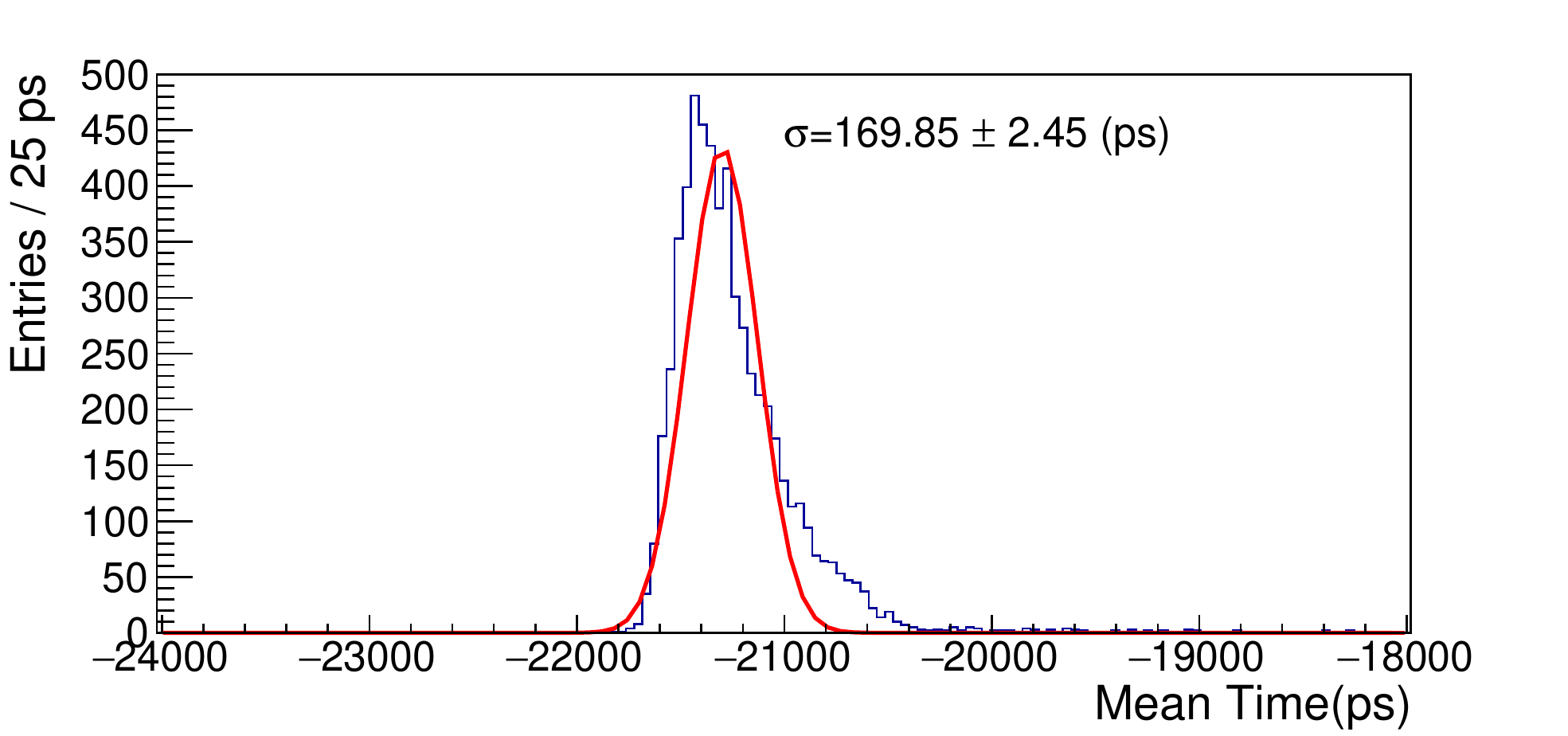}
\caption{Measured mean time distribution for the data taken at 19\,kV, 1.3\,kHz/cm$^{2}$ and with 100\% $\ECOgas$.}
\label{fig:uncorrEco1300}
\centering\includegraphics[width=0.5\textwidth]{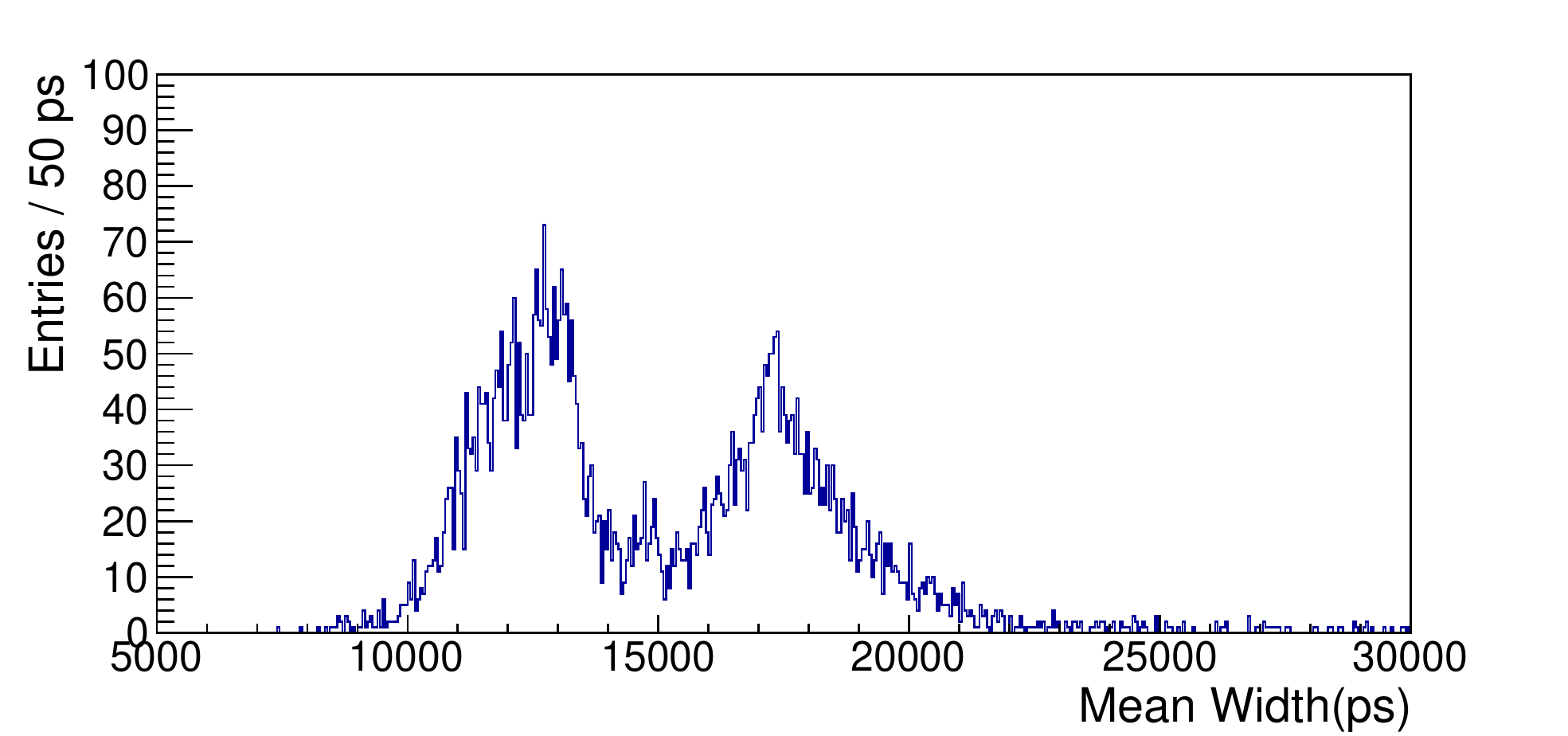}
\caption{Time-over-threshold (average of two ends) 100\% $\ECOgas$ distribution for the data taken at 19\,kV, 1.3\,kHz/cm$^{2}$.}
\label{fig:MWidthECO1300}
\centering\includegraphics[width=0.5\textwidth]{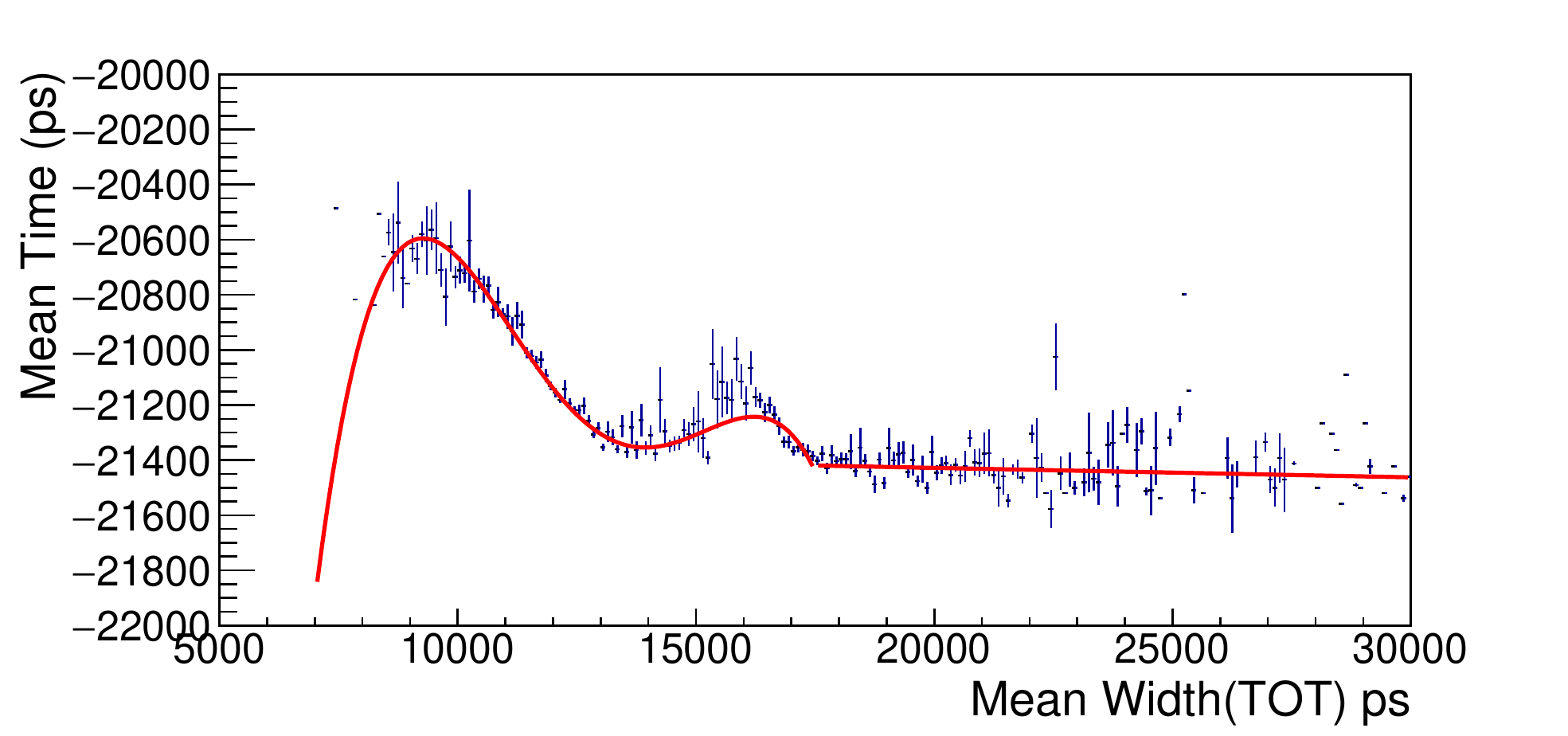}
\caption{Profile of time vs. time-over-threshold 100\% $\ECOgas$ and a fourth and a first order polynomial fit function lines for the time-slewing correction are shown. }
\label{fig:TApaper.pdf}
\centering\includegraphics[width=0.5\textwidth]{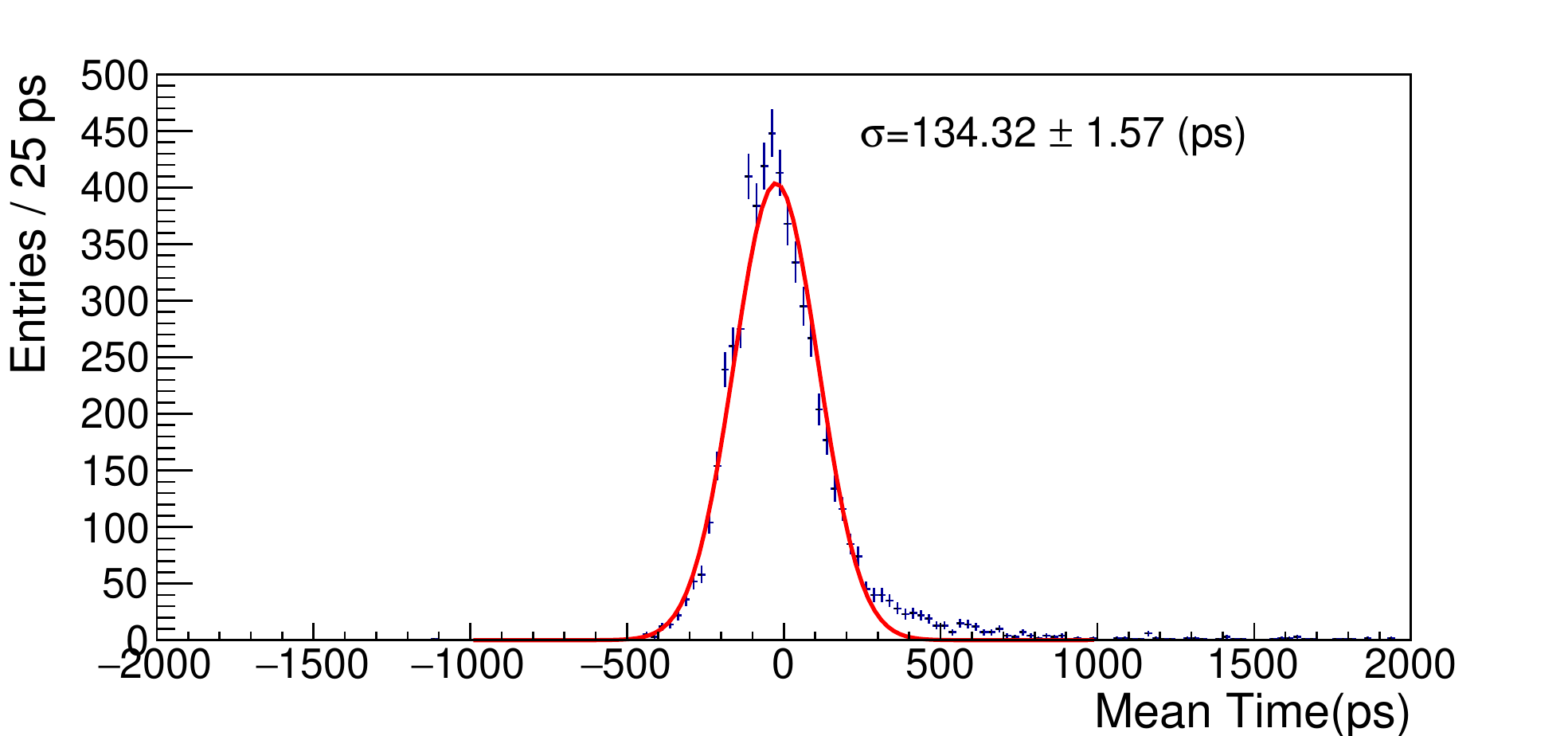}
\caption{Time-slewing corrected mean time distribution for the 100\% $\ECOgas$. }
\label{fig:corrEco1300}
\end{figure}

To obtain position independent time resolution, the mean time of signals at both ends is used to calculate the time resolution.  The measured mean time distribution and the mean width distribution for the pure $\ECOgas$ is shown in figure~\ref{fig:uncorrEco1300} and~\ref{fig:MWidthECO1300}.   Using the time-of-threshold technique (TOT) of NINO chip can derive a time-slewing correction that depends on an amount of charge. To correct this time-slewing effect a fourth  order polynomial fit function is used. 
\begin{figure}[!t]
\centering\includegraphics[width=0.5\textwidth]{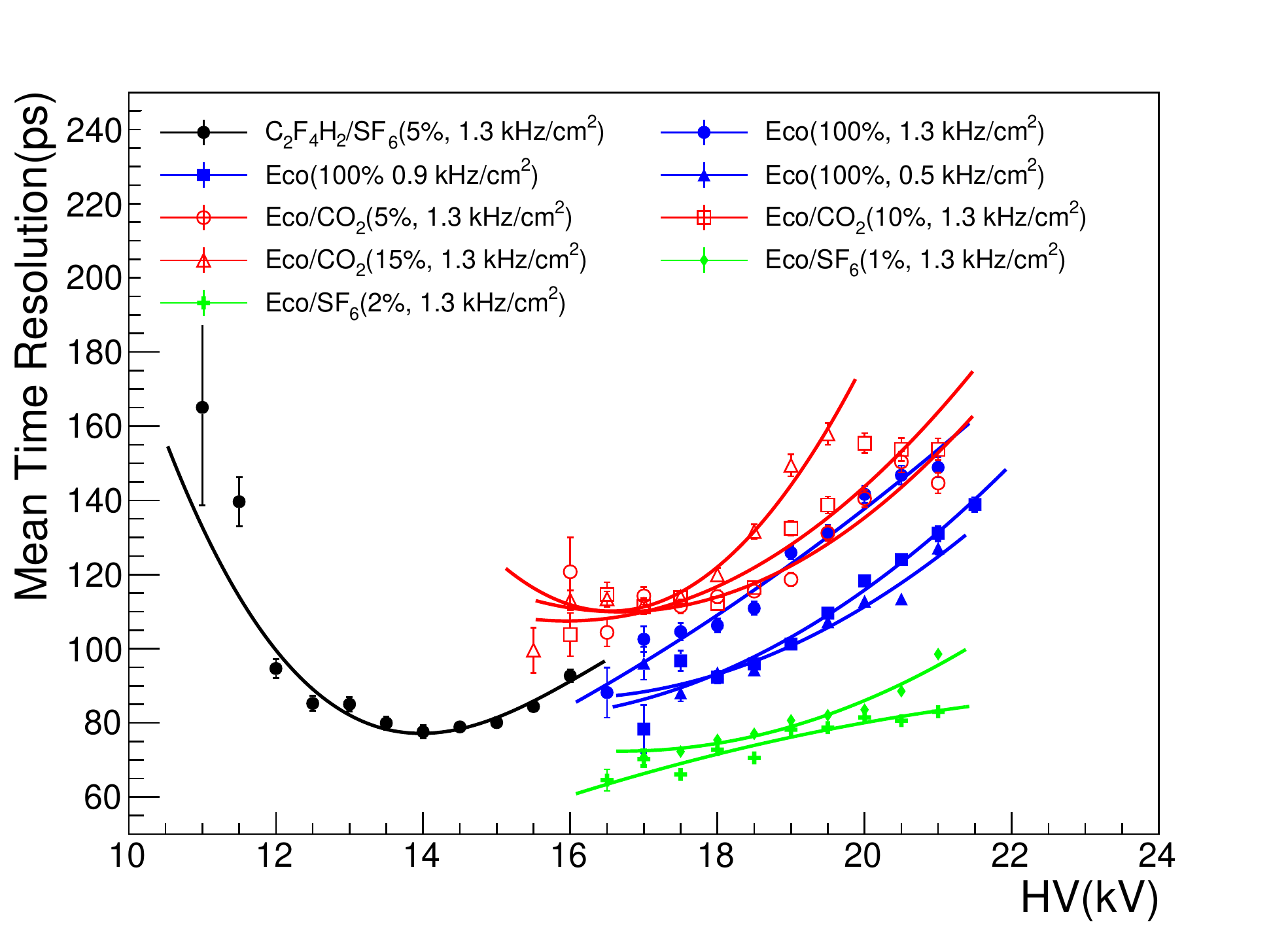}
\caption{Time resolutions of various gas mixtures at 1.3\,kHz/cm$^2$ and of different particle flux on $\ECOgas$.}
\label{fig:MTR1300}
\centering\includegraphics[width=0.5\textwidth]{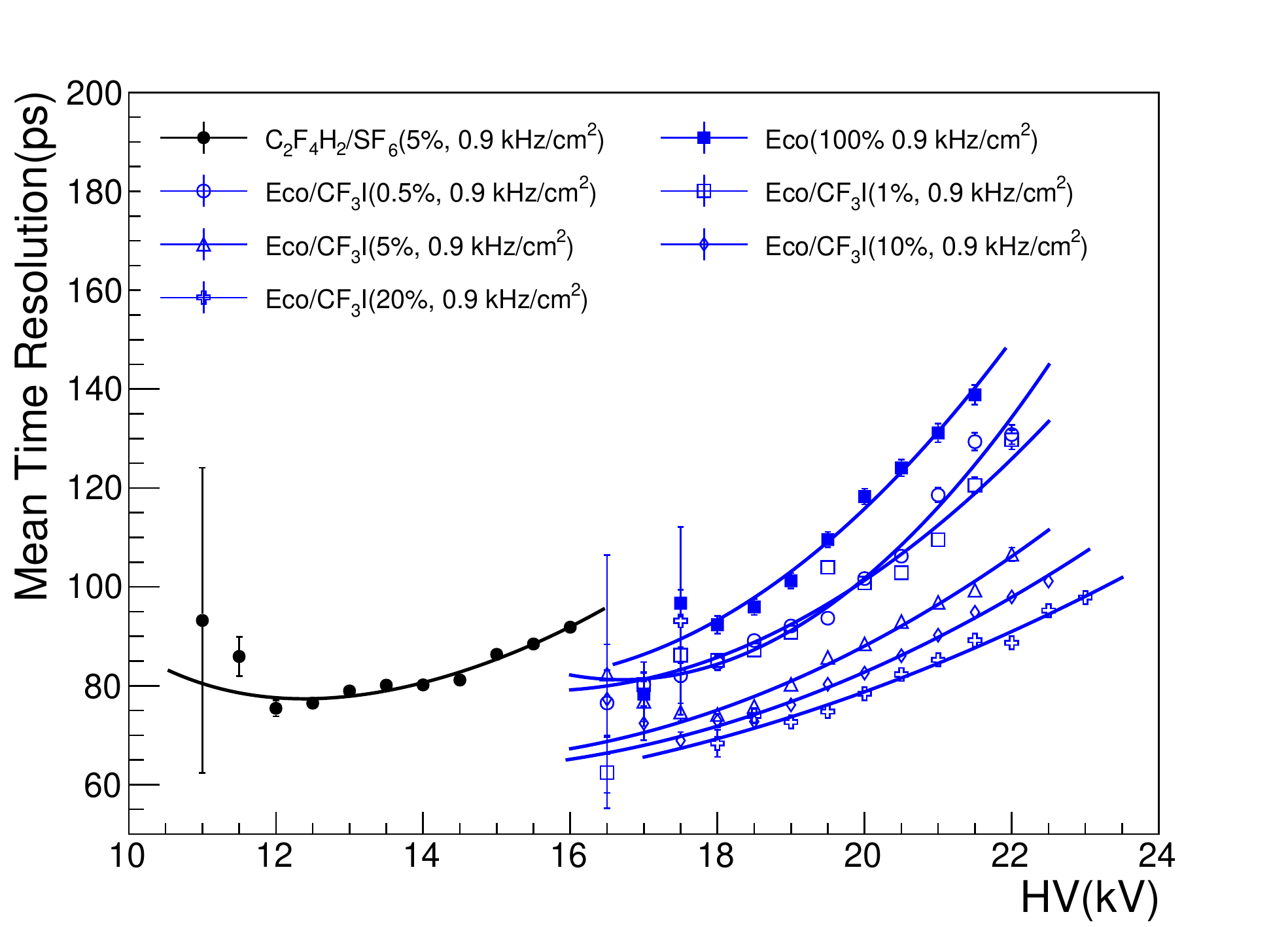}
\caption{Time resolutions of various gas mixtures at 0.9\,kHz/cm$^2$.}
\label{fig:MTR900}
\end{figure}

The measured time resolutions for different gas mixtures are shown in figure~\ref{fig:MTR1300} and~\ref{fig:MTR900}.  
At the knee voltage of the efficiency plateau the time resolution is 88\,ps for $\STDgas/\SF6$ and $\ECOgas/\SF6$.  Others are between 115 and 125\,ps at 1.3\,kHz/cm$^2$.  A better resolution is obtained for the  $\ECOgas$, which is between 95 and 120\,ps depending on the ratio of $\CF3I$.

\subsection{Position resolution}
\begin{figure}[!t]
\centering\includegraphics[width=0.5\textwidth]{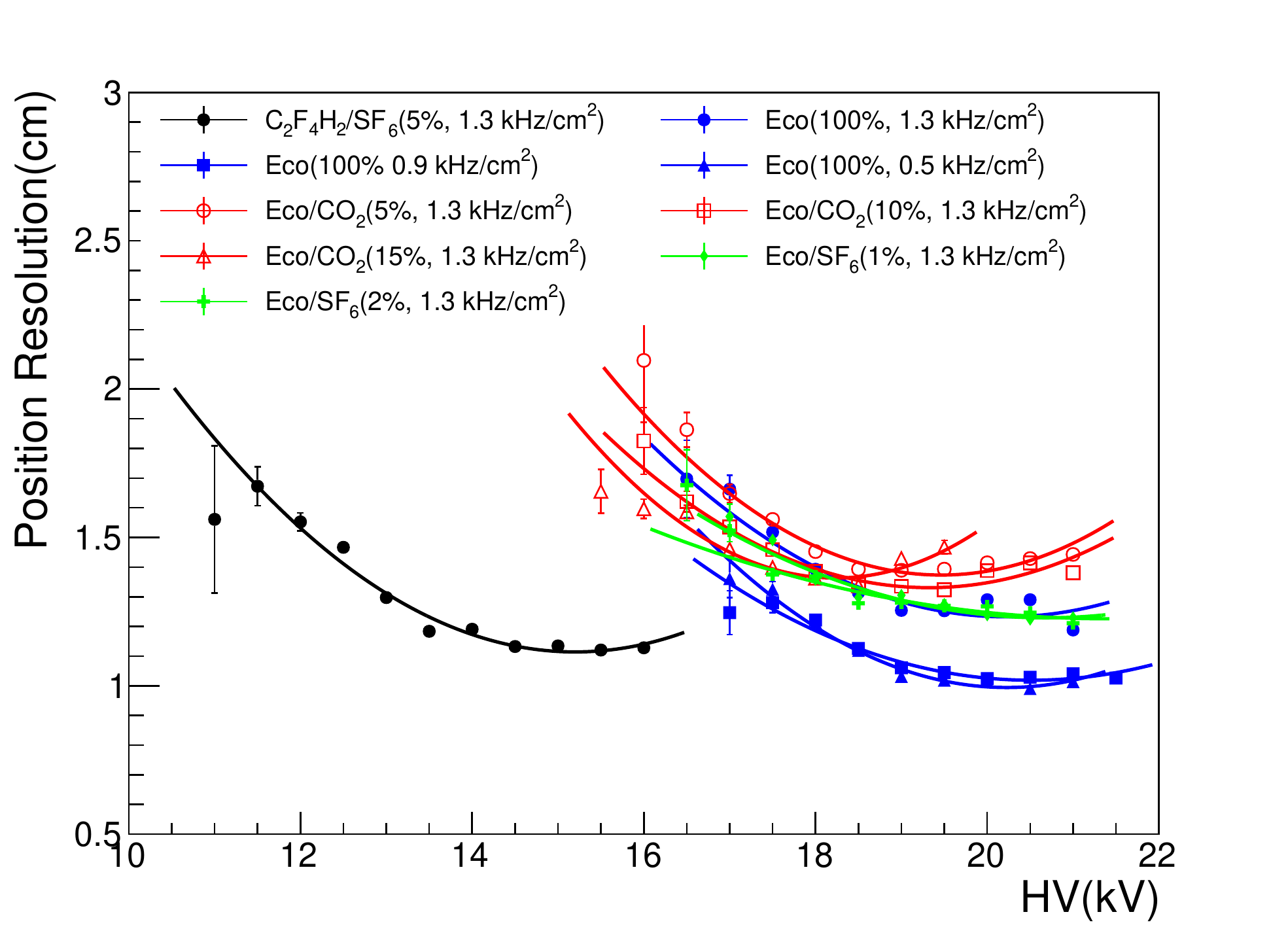}
\caption{Position resolution at 1.3 kHz/cm$^2$.}
\label{fig:position1300}
\centering\includegraphics[width=0.5\textwidth]{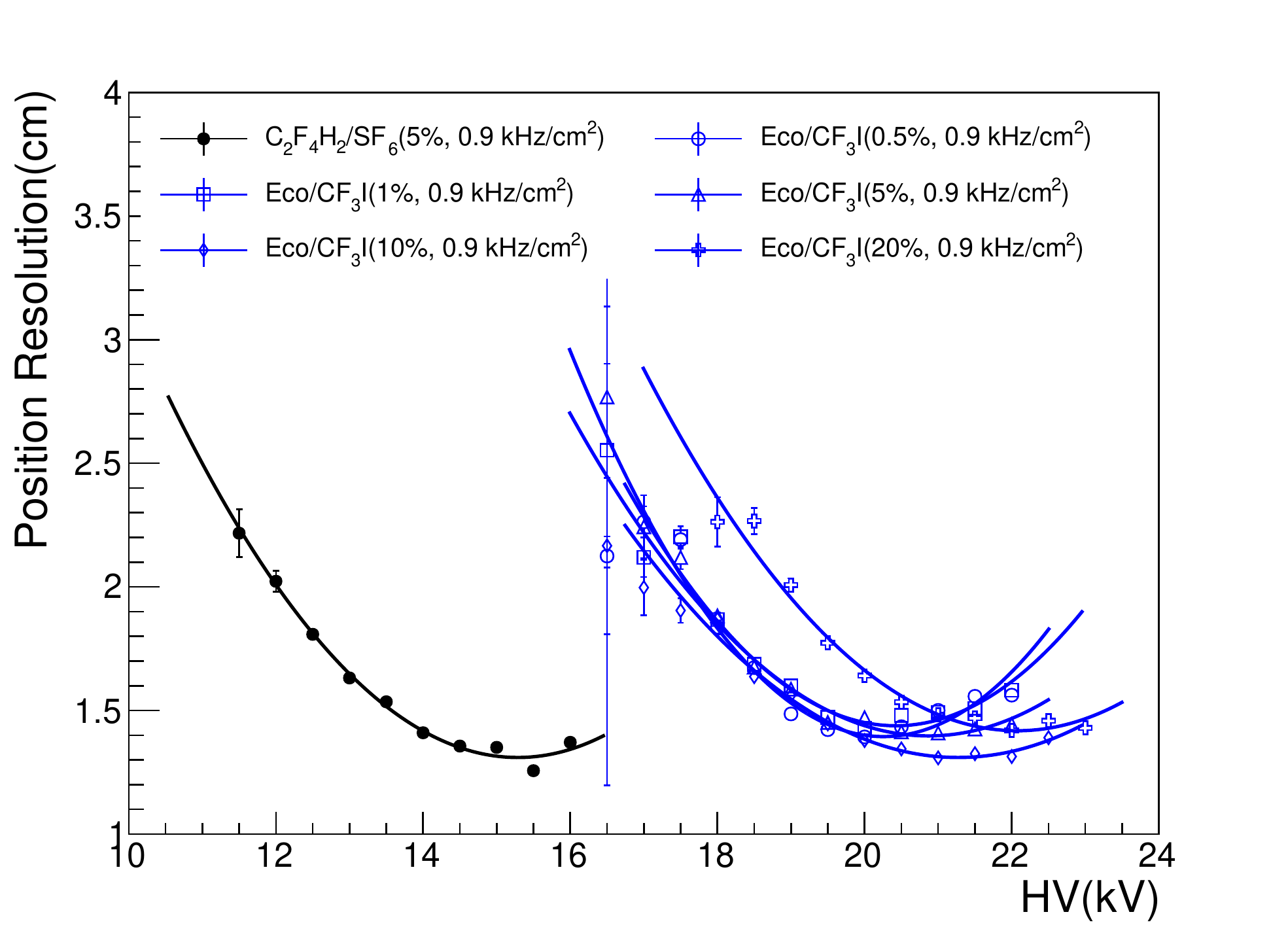}
\caption{Position resolution at 0.9 kHz/cm$^2$.}
\label{fig:position900}
\end{figure}
The position resolution can be estimated from the difference in time measured at both ends of a hit strip.  The position resolution of all the gas mixtures of $\ECOgas$ are similar to $\STDgas/\SF6$ in the same particle flux, shown in figure~\ref{fig:position1300} and~\ref{fig:position900}.   At the lower particle flux, it is improved to 1.1\,cm.

\subsection{Streamer probability}
It is difficult to define and count the number of streamers directly.  
In this analysis, a signal firing more than 5 neighbouring strips is used to define streamer event. The ratio of streamer events over the triggered events is defined as the streamer probability. 
It is assumed that the high value of streamer probability will accelerate ageing.  

\begin{figure}[h]
\centering\includegraphics[width=0.5\textwidth]{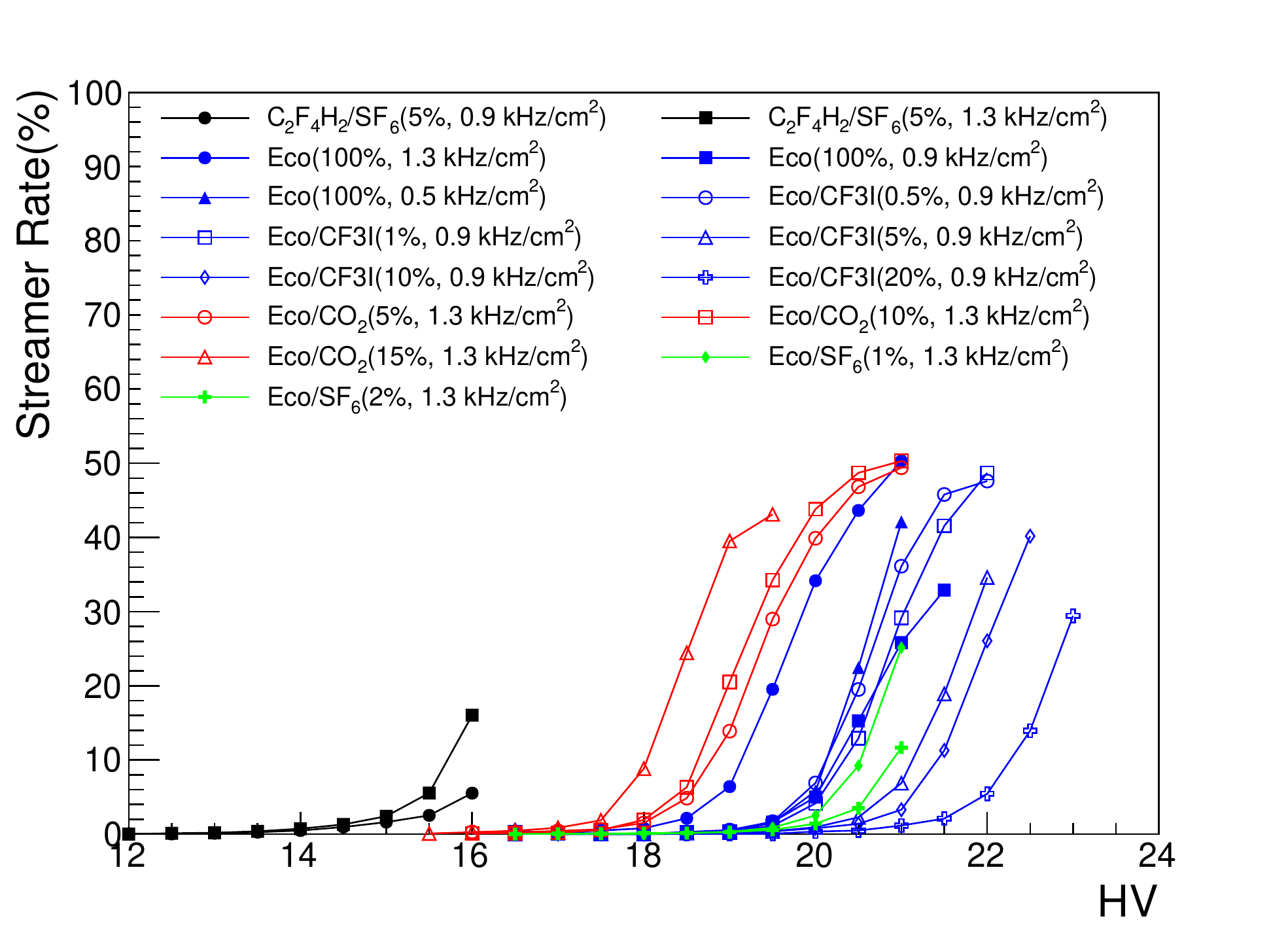}
\caption{Streamer probability for gas mixtures.}
\label{fig:streamers}
\end{figure}
The observed probability is shown in figure~\ref{fig:streamers}.
Comparing values at their operating voltages and~\ref{fig:efficiency900}), most gases have  streamer probabilities less than 5\%, except for the gas mixtures of $\ECOgas/\CO2$, which increased its value proportional to the amount of $\CO2$.   Adding $\CF3I$ to  $\ECOgas$ shows an tendency of increase of the operating voltage. It suppresses the streamer production at the knee of the efficiency plateau when increasing its ratio. To have the effective suppression of streamer, a significant amount of $\CF3I$ is needed unlike the case of adding $\SF6$.

\section{Conclusion}   

The feasibility of using the ecological gas (HFO-1234ze: $\ECOgas$, GWP $<$ 7) and also the mixture with  $\CF3I$ (GPW $<$ 1)  have been tested using a small MRPC with 6 gaps of 220\,$\mu$m, for the purpose of finding ecological gases to substitute for greenhouse gases, namely $\STDgas$ and $\SF6$, being currently used in many experiments.

For the performance of MRPC with ecological gas mixture, it needs an operating voltage of 25\% higher to reach the plateau  compared with the more commonly used gas mixture, $\STDgas/\SF6$.   Adding $\SF6$ to $\ECOgas$, the result seems almost same as the commonly used one, except for the operating voltage.    Adding $\CF3I$ to the $\ECOgas$, the efficiency plateaux becomes better depending on the amount. However, the operating voltage slightly increases. 

Overall performance of $\ECOgas/\CF3I$ (80/20\%) mixture shows a very similar result as the one of $\STDgas/\SF6$.  It should be noted here that the price of $\CF3I$ gas is currently very expensive.

\section*{Acknowledgements}
This work has been supported by the Korea-EU cooperation program of National Research Foundation of Korea, Grant Agreement 2014\-K1\-A3\-A7\-A03075053. The results presented here were obtained at the T10 test beam in the east hall at CERN. The authors acknowledge the support received by the operators of the PS.

\bibliographystyle{ieeetran}
\bibliography{references}

\begin{thebibliography}{1}
\providecommand{\url}[1]{#1}
\csname url@samestyle\endcsname
\providecommand{\newblock}{\relax}
\providecommand{\bibinfo}[2]{#2}
\providecommand{\BIBentrySTDinterwordspacing}{\spaceskip=0pt\relax}
\providecommand{\BIBentryALTinterwordstretchfactor}{4}
\providecommand{\BIBentryALTinterwordspacing}{\spaceskip=\fontdimen2\font plus
\BIBentryALTinterwordstretchfactor\fontdimen3\font minus
  \fontdimen4\font\relax}
\providecommand{\BIBforeignlanguage}[2]{{%
\expandafter\ifx\csname l@#1\endcsname\relax
\typeout{** WARNING: IEEEtran.bst: No hyphenation pattern has been}%
\typeout{** loaded for the language `#1'. Using the pattern for}%
\typeout{** the default language instead.}%
\else
\language=\csname l@#1\endcsname
\fi
#2}}
\providecommand{\BIBdecl}{\relax}
\BIBdecl

\bibitem{eco:helium}
M.~Abbrescia \emph{et~al.}, ``Eco-friendly gas mixtures for resistive plate
  chambers based on tetrafluoropropene and helium,'' \emph{JINST}, vol.~11, p.
  P08019, 2016.

\bibitem{eco:RPC}
R.~Guida \emph{et~al.}, ``Characterization of rpc operation with new
  environmental friendly mixtures for lhc application and beyond,''
  \emph{JIST}, vol.~11, p. C07016, 2016.

\bibitem{FORSTER2016182}
R.~Forster, O.~M. Rodr\'iguez, W.~Park, A.~R. Rodr\'iguez, M.~Williams,
  A.~Zichichi, and R.~Zuyeuski, ``Study of the counting rate capability of mrpc
  detectors built with soda lime glass,'' \emph{Nucl. Instr. and Meth. A}, vol.
  830, pp. 182 -- 190, 2016.

\bibitem{Anghinolfi:818554}
F.~Anghinolfi, P.~Jarron, A.~N. Martemyanov, E.~Usenko, H.~Wenninger, M.~C.~S.
  Williams, and A.~Zichichi, ``{NINO: An ultra-fast and low-power front-end
  amplifier/discriminator ASIC designed for the multigap resistive plate
  chamber},'' \emph{Nucl. Instrum. Methods Phys. Res., A}, vol. 533, pp.
  183--187, 2004.

\bibitem{Aamodt:2008zz}
K.~Aamodt \emph{et~al.}, ``{The ALICE experiment at the CERN LHC},''
  \emph{JINST}, vol.~3, p. S08002, 2008.

\end{thebibliography}







\end{document}